\documentclass[12pt]{iopart}
\usepackage[bookmarks=true, colorlinks=true, linkcolor=blue, urlcolor=blue, citecolor=blue]{hyperref}

\usepackage{iopams}
\usepackage{graphicx}
\usepackage{makecell}
\usepackage{multirow}

\usepackage{color}
\expandafter\let\csname equation*\endcsname\relax
\expandafter\let\csname endequation*\endcsname\relax
\usepackage{amsmath}
\usepackage{tikz}

\newcommand{\orangecircle}{\tikz[baseline=-0.7ex] \draw[black,fill={rgb,255:red,255;green,153;blue,102}] (0,0) circle (1ex);}

\newcommand{\bluecircle}{\tikz[baseline=-0.7ex] \draw[black,fill={rgb,255:red,78;green,179;blue,207}] (0,0) circle (1ex);}

\newcommand{\yellowcircle}{\tikz[baseline=-0.7ex] \draw[black,fill={rgb,255:red,255;green,255;blue,0}] (0,0) circle (1ex);}

\newcommand{\purplecircle}{\tikz[baseline=-0.7ex] \draw[black,fill={rgb,255:red,112;green,48;blue,160}] (0,0) circle (1ex);}

\newcommand{\greencircle}{\tikz[baseline=-0.7ex] \draw[black,fill={rgb,255:red,99;green,165;blue,55}] (0,0) circle (1ex);}

\begin{document}

\title[Applicability of MBQC towards Physically-driven Variational Quantum Eigensolver]{Applicability of Measurement-based Quantum Computation towards Physically-driven Variational Quantum Eigensolver}

\author{Zheng Qin$^1$, Xiufan Li$^2$, Yang Zhou$^{1,3,*}$, Shikun Zhang$^1$, Rui Li$^1$, Chunxiao Du$^1$ and Zhisong Xiao$^{1,4}$}

\address{$^1$School of Physics, Beihang University, Beijing 100191, China}
\address{$^2$Centre for Quantum Technologies, National University of Singapore, 3 Science Drive 2, 117543 Singapore}
\address{$^3$Research Institute for Frontier Science, Beihang University, Beijing 100191, China}
\address{$^4$School of Instrument Science and Opto-Electronics Engineering, Beijing Information Science and Technology University, Beijing 100192, China}

\ead{yangzhou9103@buaa.edu.cn}
\vspace{10pt}

\date{\today}

\begin{abstract}
Variational quantum algorithms are considered one of the most promising methods for obtaining near-term quantum advantages; however, most of these algorithms are only expressed in the conventional quantum circuit scheme. The roadblock to developing quantum algorithms with the measurement-based quantum computation (MBQC) scheme is resource cost. Recently, we discovered that the realization of multi-qubit rotation operations only requires a constant number of single-qubit measurements with the MBQC scheme, providing a potential advantage in terms of resource cost. The structure of the Hamiltonian variational ansatz (HVA) aligns well with this characteristic. Thus, we propose an efficient measurement-based quantum algorithm for quantum many-body system simulation tasks, called measurement-based Hamiltonian variational ansatz (MBHVA). We then demonstrate its effectiveness, efficiency, and advantages with the two-dimensional Heisenberg model and the Fermi-Hubbard chain. Numerical experiments show that MBHVA can have similar performance as circuit-based ansatz, and is expected to reduce operation counts during execution compared to quantum circuits, bringing the advantage of running time. We conclude that the MBQC scheme is potentially feasible for achieving near-term quantum advantages in the noisy intermediate-scale quantum (NISQ) era, especially in the presence of large multi-qubit rotation operations.
\end{abstract}

%
\vspace{2pc}
\submitto{\NJP}
%
%

\section{Introduction}
\label{section:I}

\ \ \ \ In recent years, quantum computation has emerged as a promising approach for solving a variety of classically intractable problems, such as simulating complicated quantum systems or solving large-scale linear algebra problems. Constrained by the number of qubits and fidelity of quantum operations, current quantum hardware is considered to be in the noisy intermediate-scale quantum (NISQ) regime \cite{NISQ}. This has led to an interest in hybrid quantum-classical algorithms \cite{hybrid-qc}, which embed the problems into short-depth parameterized quantum circuits, particularly suited to NISQ hardware, and employ classical optimization routines to find the quantum circuits that best solve the problem at hand. Of these, the most promising is the variational quantum algorithm (VQA), which is arguably the quantum analog of highly successful machine-learning methods \cite{VQA}. Famous examples of VQA include the variational quantum eigensolver (VQE) \cite{vqe} and the quantum approximate optimization algorithm (QAOA) \cite{QAOA1, QAOA2}. These algorithms have been considered for a plethora of applications, ranging from foundational research to engineering practice. Existing hybrid quantum-classical algorithms are primarily theoretically formulated in circuit-based quantum computation (CBQC). Although they may be key to obtaining near-term quantum advantages, there are still many important challenges, including their trainability, accuracy, efficiency, and compatibility with photonic platforms \cite{vqe_review1, vqe_review2, barren_plateaus, barren_plateaus_new, vqa_np}. 

Measurement-based quantum computation (MBQC) is another general quantum computation scheme. In contrast to CBQC which manipulates the evolution of qubits using quantum gates to realize specific computation, MBQC mainly uses resource states and measurement operations to accomplish computational tasks. The single-measurement procedure can be of high fidelity and robust to environmental noise. More importantly, measurement-based VQE (MBVQE) is suitable for achieving quantum advantages on photonic platforms \cite{tilly2022variational, arzani2021harmonizing, kashif2023physical}. At present, there are two main approaches for realizing the variational quantum computation process under MBQC. The first approach is to construct the required graph state and measurement pattern by translating from circuit-based VQE (CBVQE). This method is simple to implement but may result in high resource costs that the current hardware cannot afford. Here, resource costs refer to the number of operations and qubits required in the MBQC scheme. Fortunately, several reduction schemes exist for optimizing the graph state~\cite{MBQPE, MBHEA}. However, not all VQE ans$\rm \ddot{a}$tze are suitable for translation. Both the size of the graph state and measurement patterns are affected by the VQE ansatz. As a result, even if two converted graph states are equivalent, their resource costs may differ. The second approach is to construct the so-called `custom state' (or native graph state) and `customized variational knobs' (or native rotation operations) according to the specific problems. Recent research~\cite{MBQAOA} found that resonating natively, compared with translating circuit-based algorithms into MBQC, can significantly reduce the operation counts and qubits required. However, this method faces significant difficulties in the direct construction of graph states.

Despite the difficulties of graph state constructions, MBQC theoretically offers a primary advantage of the equivalent execution of complex quantum gates that are challenging to implement with high fidelity. The evolution process of MBQC is driven by consecutive single-qubit measurements, which implies that the operations of complex quantum gates in CBQC can be converted into a series of high-fidelity single-qubit measurements. For instance, multi-qubit rotation operations that have a wide range of applications in quantum simulation and algorithms adversely affect computational accuracy owing to the low fidelity of the current physical hardware. However, in MBQC, they can be realized by using multiple high-fidelity single-qubit measurement operations and may lead to lower resource costs. In this work, we will clarify this point with the proposed measurement-based Hamiltonian variational ansatz (MBHVA). With the help of Hamiltonian variational ansatz (HVA), we can formulate the mathematical expressions that represent the unitary operations the ansatz aims to implement and construct a `native graph state' according to the specific form of the Hamiltonian, thus simultaneously avoiding the difficulties of direct construction and optimization of the graph state. Numerical simulations with two paradigmatic many-body systems have demonstrated that its computational performance is similar to circuit-based Hamiltonian variational ansatz (CBHVA) and superior to both measurement-based hardware efficient ansatz (MBHEA) and circuit-based hardware efficient ansatz (CBHEA). Furthermore, with the presence of large multi-qubit rotation operations which is common in physically-driven ansatz, MBHVA has less operation count than CBHVA during execution which may lead to the advantage in running time.

The remainder of this paper is organized as follows. In Section \ref{section:II}, we introduce the basic concepts of VQE and HVA, as well as MBQC. In Section \ref{section:III}, we demonstrate the proposed MBHVA and verify its effectiveness and efficiency using two quantum many-body models: the two-dimensional Heisenberg model and the Fermi-Hubbard chain. This section presents the main results. Finally, in Section \ref{section:IV}, we summarize the paper and discuss the issues of MBVQE.

\section{Theoretical background}
\label{section:II}

\subsection{Variational quantum eigensolver and Hamiltonian variational ansatz}
\label{section:2.1}

\ \ \ \ Variational Quantum Eigensolver (VQE) aims to compute the ground state energy of a Hamiltonian, which is generally the first step in studying the energetic properties of molecules and materials. The algorithm starts by representing the Hamiltonian by a quantum circuit characterized by two parts: a set of ordered quantum gates often referred to as an `ansatz' and a set of parameters that determines the operation of these quantum gates. The number of consequential operations in a circuit is referred to as the depth. The state of the qubits is designed to model a trial wave function, and the studied Hamiltonian can be measured with respect to this wave function to estimate the ground state energy. Subsequently, a classical optimization routine is applied to determine the optimal parameters for the quantum circuit, which minimizes the expectation value of the Hamiltonian under the variational principle. This process was repeated iteratively until convergence was achieved.

The design of the ansatz is at the core of the potential advantage of VQE over classical methods. The Hamiltonian variational ansatz (HVA) \cite{HVA} is aimed at constructing a series of quantum gate operations based on the specific Hamiltonian. The first step of constructing a quantum circuit with HVA is to decompose the Hamiltonian into a series of local Hamiltonians using Trotter decomposition \cite{trotter} as follows:

\begin{equation}
{
 H = \sum_{a}H_{a},
}
\label{eq:HVA-operators}
\end{equation}
where local Hamiltonians $H_{a}$ does not commute and is often composed of the tensor-product of the Pauli operators $\sigma=\{I, X, Y, Z\}$. For the construction of $D$ layers, also called the depth of the ansatz, HVA can be expressed as

\begin{equation}
{
 |\psi_{D}\rangle = \prod^{D}_{d=1} \left(\prod_{a} {\rm exp}(-i\theta_{a,d}H_{a}) \right)|\psi_{0}\rangle,
}
\label{eq:HVA-layers}
\end{equation}
where $|\psi_{0}\rangle$ is the ground state of one of the terms, and $\theta_{a,d}$ are the rotation angles, which are also the parameters. 

HVA has been proven particularly suitable for determining the ground state energy of quantum many-body systems, and it demonstrates strong resilience to the barren plateau (BP) problem \cite{barren_plateaus, HVA-practice, park2024hamiltonian, holmes2022connecting, mele2022avoiding}. We used the Transverse Field Ising Model (TFIM) to illustrate the HVA construction process. TFIM is a quantum spin model with two-body interactions between nearest-neighbor spins. For simplicity, we selected an open boundary as the boundary condition. The Hamiltonian for the one-dimensional TFIM is
\begin{equation}
{
 H_{TFIM} = -\sum^{N}_{i=1}(J\cdot Z_{i}Z_{i+1} + \Gamma \cdot X_{i}) = J\cdot H_{ZZ} + \Gamma \cdot H_{X},
}
\label{eq:HVA-1DTFIM}
\end{equation}
where $N$ is the total number of spins, $J$ represents the strength of the spin-spin interaction, $\Gamma$ is the transverse field strength. And $X_{i}$ ($Z_{i}$) donates the corresponding Pauli operator $X$ ($Z$) that acts on the $i-th$ particle.

According to Eqs.~(\ref{eq:HVA-layers}) and (\ref{eq:HVA-1DTFIM}), we can represent the unitary operation $U$ of the quantum circuit as
\begin{equation}
{
 U_{TFIM}(\beta, \gamma) = \prod^{D}_{d=1}{\rm exp} \left(-i \frac{\beta_{d}}{2}H_{ZZ} \right) {\rm exp} \left(-i \frac{\gamma_{d}}{2}H_{X} \right),
}
\label{eq:HVA-U}
\end{equation}
where $\beta_{d}$ and $\gamma_{d}$ are the parameter vectors correspond to layer $d$.

It is easy to see that this quantum circuit established by HVA mainly consists of single-qubit rotation gates $R_{X}(\theta)={\rm exp}\left(-i\theta X / 2 \right)$ and two-qubit rotation gates $R_{ZZ}(\theta)={\rm exp}\left(-i\theta Z \otimes Z / 2 \right)$. In general, quantum many-body systems with more complex Hamiltonian forms result in more complex rotation operations. In the MBQC scheme, these rotation operations, especially the multi-qubit rotation $Z$ operations, may have a more convenient physical implementation than quantum circuits. This is discussed in Section \ref{section:2.3}.

It is worth mentioning that the depth of the VQE affects the expressivity and entanglement capability of the quantum circuit \cite{Expressibility, Expressivity}. Increasing the number of layers may improve the performance. However, when implementing VQE on quantum hardware, large circuit layers cause error accumulation and deviate from the theoretical calculations \cite{vqe_review1}. Therefore, it is important to select the appropriate depth. For one-dimensional TFIM, Ref.~\cite{HVA-practice, HVA-depth, variational-simulation} suggest that HVA can exactly reach the ground state energy at $D=N/2$, where N is the number of qubits.

\subsection{Measurement-based quantum computation}
\label{section:2.2}

\ \ \ \ Measurement-based quantum computation (MBQC) is an alternative to the quantum circuit model and its basic idea is that once a certain resource state is available, the universal quantum computation can be executed by local projective measurements \cite{MBQC, MBQC_review}. There are two highly representative theories: teleportation-based quantum computation (TQC) \cite{TQC} and one-way quantum computer (1WQC) \cite{1WQC}. In the following, we take 1WQC as an example to introduce MBQC in detail. 

\subsubsection{Primary processes of MBQC.}

For MBQC, a multi-qubit entangled state should be prepared first as resource state. A common type of resource state has a structure consisting of vertices and edges just as a graph and is therefore called graph state \cite{LC-equivalence, graph_state}. Its corresponding mathematical expression is

\begin{equation}
    |G\rangle=\prod\limits_{(v_{i},v_{j})\in E}CZ_{v_{i},v_{j}}(\bigotimes_{v_{k} \in V}|+\rangle_{v_{k}})
\label{eq:MBQC-graph_state}
\end{equation}
where $G$ represents the graph, $|G\rangle$ is the generated graph state, $V$ is the set of vertices, $E$ is the set of edges, $CZ$ is the controlled-$Z$ gate \cite{MBQC_resource}. $|\pm\rangle=\frac{1}{\sqrt{2}}(|0\rangle\pm|1\rangle)$ are the eigenstates of the Pauli $X$ operator. Both $|+\rangle$ and $|-\rangle$ states can serve as single-qubit states. In this work, we use $|+\rangle$ without distinction.

Once the resource state is prepared, the next step is to perform local projective measurements in a given sequence. In MBQC, single-qubit measurements can be performed by choosing different projection planes. The measurement basis is represented as $\{|\uparrow(\theta, \varphi)\rangle, |\downarrow(\theta, \varphi)\rangle\}$, specified in Eqs.~(\ref{eq:MBQC-basis1}) \& (\ref{eq:MBQC-basis2}) \cite{1WQComputation, Hybrid-quantum-computation}. They can also be abbreviated as the measurement angles $(\theta, \varphi)$ in the Bloch sphere.
\begin{eqnarray}
|\uparrow(\theta, \varphi)\rangle &=& {\rm cos}\left(\frac{\theta}{2}\right)|0\rangle+e^{i\varphi}{\rm sin}\left(\frac{\theta}{2}\right)|1\rangle
, \label{eq:MBQC-basis1}
\\
|\downarrow(\theta, \varphi)\rangle &=& -{\rm sin}\left(\frac{\theta}{2}\right)|0\rangle+e^{i\varphi}{\rm cos}\left(\frac{\theta}{2}\right)|1\rangle
. \label{eq:MBQC-basis2}
\end{eqnarray}

After the measurement, the quantum state undergoes by-product evolution involving Pauli operations. These operators in by-product evolution are called by-product operators. To complete the expected evolution process, the final step is to correct the effects of these by-products. The specific correction process is determined by the measurement outcomes and thus is known as outcome dependence.

For common rotation operations in variational algorithms, the byproduct operators are Pauli operators, which can be swapped with Pauli measurements based on commutation and anti-commutation relations. Thus, these byproduct operators can be propagated and combined at the end of the pattern according to the CME technique \cite{danos2007measurement}. After completing all measurements, corrections can be performed on the measurement results based on the final Pauli operators that need to be corrected in the specific output qubit. This correction process can be classical and has a small resource overhead \cite{yamasaki2020polylog}. 

In summary, the execution of MBQC tasks involves a combination of classical and quantum processes, with the classical process determining the measurement basis based on the measurement outcomes and the final correction, while the quantum process involves building the graph state and measurement operations.

\subsubsection{Preparation of graph states and measurement patterns.}

The theoretical support for determining the optimal graph state is local complementation equivalence (LC-equivalence) \cite{LC-equivalence}. If there is a set of local quantum operations that can transform one state into another, then these two multi-particle quantum states are considered to be LC-equivalent. These local operations typically include single-qubit operations, such as single-qubit quantum gates, and more general operations, such as measurements and classical feedback.

LC-equivalence is important for determining whether a given graph state can be used to perform a specific quantum algorithm. Specifically speaking, we can first propose a question-driven graph state, and then try to optimize it by practically feasible operations with the least resource costs. If such an LC-equivalence transformation can be found, the computational process of this quantum circuit can be equivalently implemented in the MBQC. For example, the Gottesmann-Knill theorem \cite{G-K-theory} shows that local Clifford operations are equivalent to the quantum states obtained after performing all non-adaptive measurements, which can effectively guide the optimization of the graph state construction \cite{G-K-theory2}. Generally, the preparation of graph states and measurement patterns consists of the following three steps.

$Step\ 1$. Creating initial graph states and corresponding measurement patterns. This is realized by starting with a set of qubits all in the $|+\rangle$ state and then applying a series of $CZ$ gates between the qubits to form multi-qubit entangled states. 

$Step\ 2$. Optimizing the graph state and its measurement pattern according to LC-equivalence. The optimization objectives are easier physical implementation, lower resource costs (measurement operations and number of qubits), and improved computational fidelity.

$Step\ 3$. Initializing the input state. This can be accomplished by measuring an additional subset of qubits in the graph state, according to the setting of the input state.

\subsection{Multi-qubit rotation operations in MBQC}
\label{section:2.3}

\begin{figure}[htb]
\centering
\includegraphics[width=0.6\linewidth]{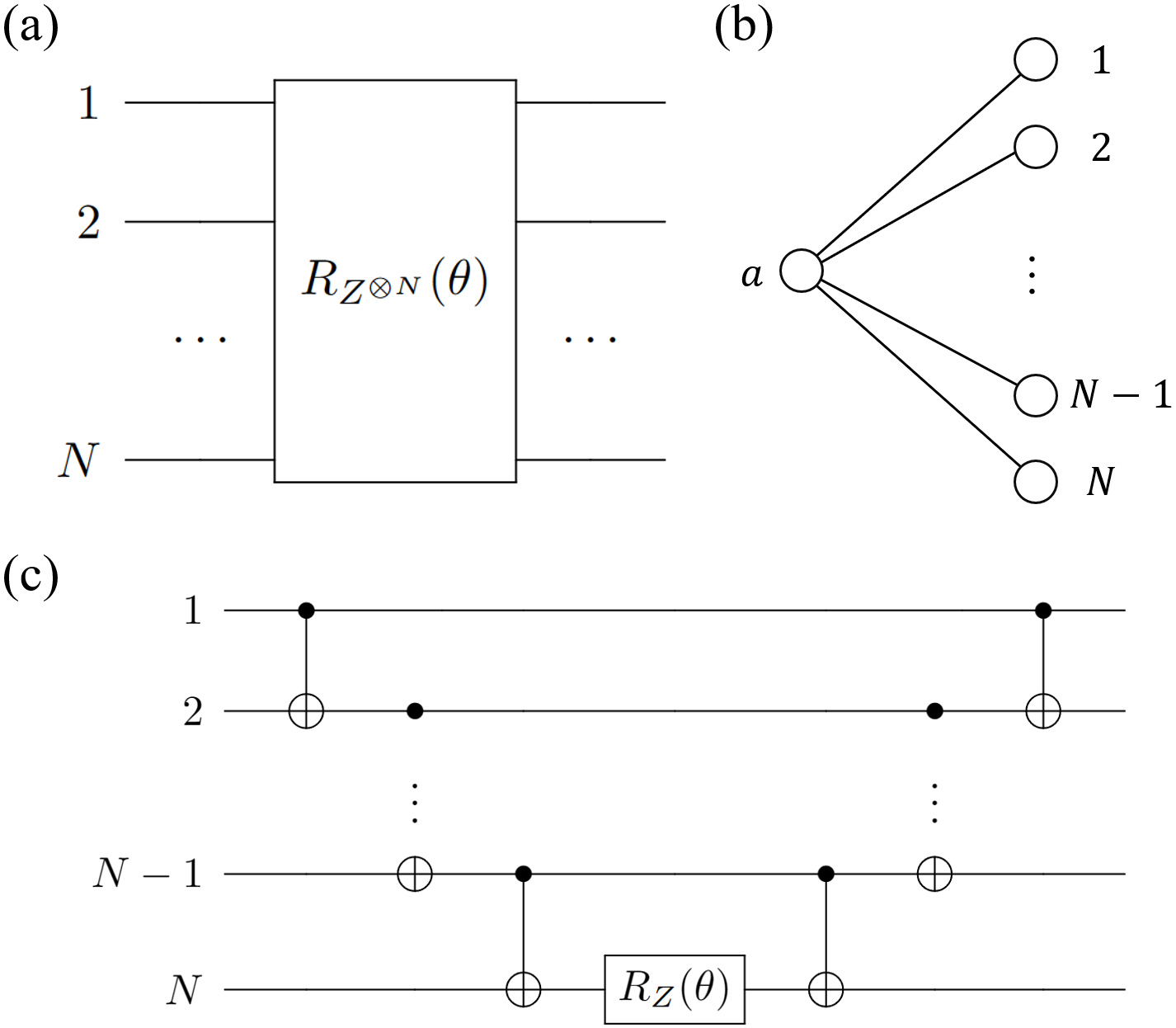}
\caption{\label{fig:multi z}{(a) A multi-qubit $Z$ rotation gate in CBQC scheme. (b) Schematic representation of the multi-qubit $Z$ rotation gate in MBQC scheme, where an additional ancillary qubit $a$ is introduced and a single measurement on it enables the equivalent operation. (c) Decomposition of the multi-qubit $Z$ rotation gate into the combination of a $R_{Z}(\theta)$ and multiple $CNOT$ gates.}}
\end{figure}

As mentioned in the previous section, a quantum circuit established by HVA mainly consists of various multi-qubit rotation gates. Among these, one of the most common is the multi-qubit rotation $Z$ gate $R_{Z^{\otimes N}}(\theta)={\rm exp}\left(-i\theta Z^{\otimes N} / 2\right)$, which can be represented in the quantum circuit as shown in Fig.~\ref{fig:multi z}(a). In the MBQC scheme, we can represent it as Fig.~\ref{fig:multi z}(b), where $a$ is the ancillary qubit, surrounded by the relevant $N$ qubits. For such a graph state, we can implement the corresponding multi-qubit rotation $Z$ gate with a single measurement operation on the ancillary qubit. The basis of this measurement operation is $(\theta, \pi/2)$.

However, the $N$-qubit state obtained by measuring this graph state is not the same as the output state $|\psi_{out}\rangle=R_{Z^{\otimes N}}(\theta)|\psi_{in}\rangle$. This is because we still need to deal with the by-product operators generated during evolution. The actual graph state evolution equation considering the by-product operators is
\begin{equation}
 |\psi_{out}\rangle=(Z^{\otimes N})^{m_{a}}U_{Z^{\otimes N}}(\theta)|\psi_{in}\rangle,
\label{eq:MBQC-Uzzz}
\end{equation}
where $U_{Z^{\otimes N}}(\theta)$ represents the evolution of measuring the ancillary qubit $a$, $m_{a}$ is the measurement outcome, $Z^{\otimes N}$ is the production of by-product operators \cite{Hybrid-quantum-computation}. There are two possible measurement outcomes, $m_{a}=0$ and $m_{a}=1$. The correction is only required when $m_{a}=1$. When $m_{a}=0$, $Z^{\otimes N}$ will degenerate into the identity operator $I$. This demonstrates that the correction should be adjusted according to specific measurement results. Refer to~\ref{appendix:A} for more details.

In addition, two other types of multi-qubit rotation gates are also common in quantum many-body systems:
\begin{eqnarray}
R_{X_{1}Z^{\otimes n}X_{N}}(\theta)&=&{\rm exp}\left(-i\theta X \otimes Z^{\otimes n} \otimes X / 2\right),
\\
R_{Y_{1}Z^{\otimes n}Y_{N}}(\theta)&=&{\rm exp}\left(-i\theta Y \otimes Z^{\otimes n} \otimes Y / 2\right),
\end{eqnarray}
where $n=N-2$, and the subscript of the Pauli operator indicates the $i^{th}$ qubit associated with multi-qubit rotation operations.

They can be realized by a multi-qubit rotation Z gate with additional single-qubit rotation gates:
\begin{equation}
   R_{X_{1}Z^{\otimes n}X_{N}}(\theta)=R_{Y_{1}}(\beta)R_{Y_{N}}(\beta)R_{Z^{\otimes N}}(\theta)R_{Y_{1}}(\gamma)R_{Y_{N}}(\gamma)
, \label{eq:Rxzx}
\end{equation}
\begin{equation}
    R_{Y_{1}Z^{\otimes n}Y_{N}}(\theta)=R_{X_{1}}(\gamma)R_{X_{N}}(\gamma)R_{Z^{\otimes N}}(\theta)R_{X_{1}}(\beta)R_{X_{N}}(\beta)
, \label{eq:Ryzy}
\end{equation}
where $\beta = -\gamma = \pi/2$.

The Hamiltonians of quantum many-body systems generally include long-range interactions which, in physically-driven ansatz, are often represented by multi-qubit rotation operators. Therefore, physically-driven VQE, particularly HVA, are applicable to the MBQC scheme. For example, the 1D-TFIM mentioned in Section \ref{section:2.1} has a two-qubit rotation $Z$ gate $R_{ZZ}$ which is particularly suitable for using this characteristic to design the computation process with the MBQC.

To explore the advantage of this characteristic of MBQC brings to the computational process, we analyze it theoretically and the following results discussed are obtained by classical simulation rather than by real quantum hardware.

\section{Measurement-based Hamiltonian variational ansatz}
\label{section:III}

\begin{figure*}[ht]
\includegraphics[width=1.0\linewidth]{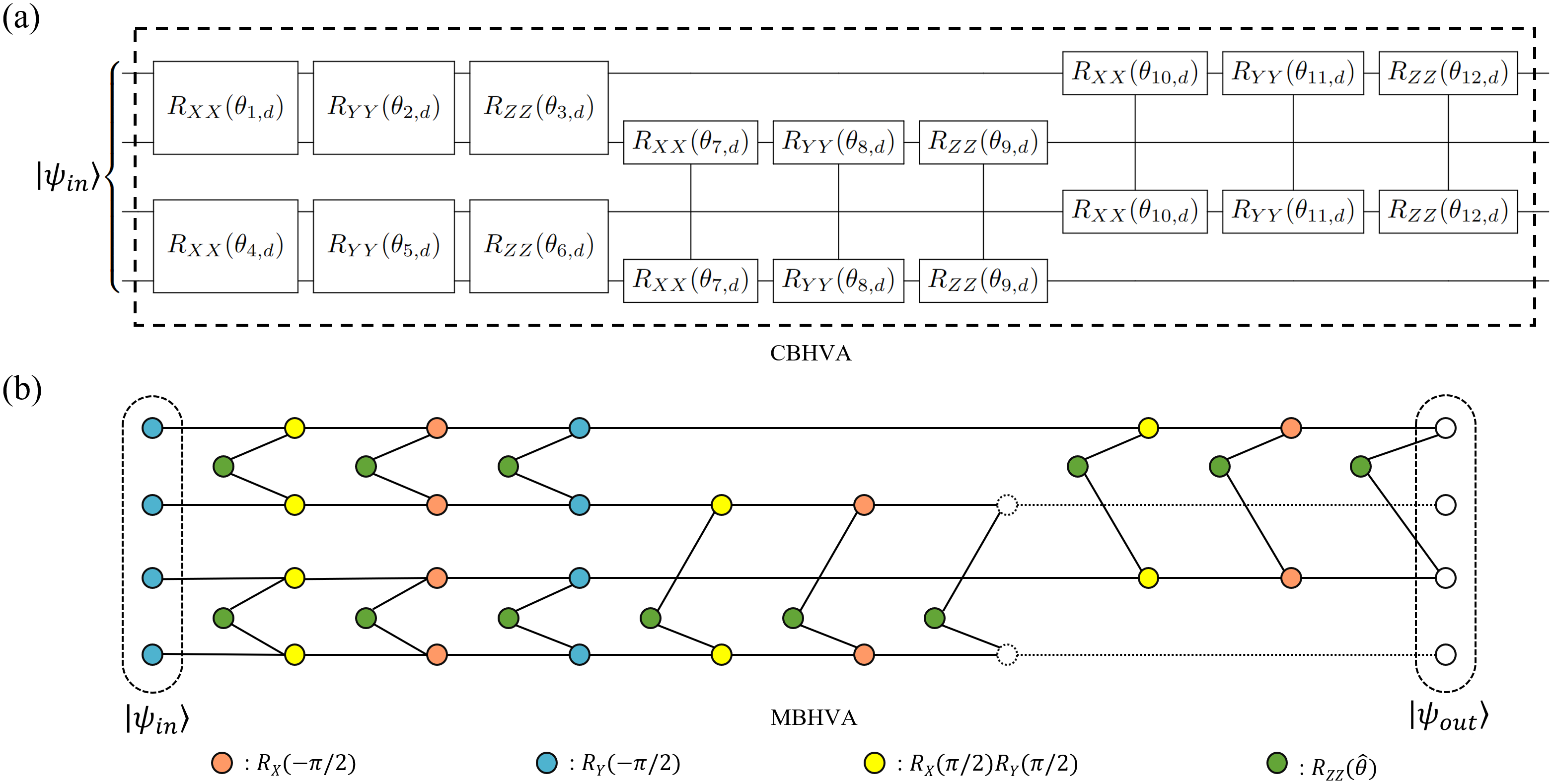}
\caption{\label{fig:MBHVA-Heisenberg graph state}(a) The quantum circuit construction of the CBHVA, where $\theta_{i,d}$ represents the rotation angle of each two-qubit rotation gate, and is also the variational parameter. The quantum circuit within the dashed box can be repeated $D$ times. (b) Schematic representation of the corresponding one-layer MBHVA graph state, where each node, except for the white output qubit, is associated with a specific operation and represented by different colors. The measurement order of the graph state is from left to right. Each solid line represents $CZ$ entanglement between the corresponding qubits, and $\hat{\theta}$ is the parameter vector. To provide a better demonstration, we introduce dashed-line nodes, referred to as `phantom nodes', which are just the virtual clone of the node on the other end of the dashed line.}
\end{figure*}

\ \ \ \ As discussed above, the quantum circuits established with HVA mainly consist of multi-qubit rotation gates whose operation can be equivalently realized by a single measurement on the ancillary qubit in the MBQC. Therefore, measurement-based Hamiltonian variational ansatz (MBHVA) can be particularly suitable for finding the ground state energy of quantum many-body systems in the NISQ era. In this section, we verify and demonstrate MBHVA with the two-dimensional Heisenberg model and the Fermi-Hubbard chain respectively. The impact of noise is not considered but is discussed in Section \ref{section:IV}. In the following, we refer to the HVA based on the quantum circuit as CBHVA (circuit-based HVA) to better distinguish between the two different schemes for construction. PaddleQuantum \cite{Paddlequantum} has been used for numerical experiments, and it is a classical simulation platform for both computational schemes. The Adam Optimizer has been selected for the optimization algorithm.

\subsection{Two-dimensional Heisenberg model}
\label{section:3.1}

\ \ \ \ The Heisenberg model is a paradigmatic model used in quantum magnetism studies. To effectively verify the MBHVA, we consider a two-dimensional square lattice with a total particle number of $N=n \times n$, where $n$ is the number of particles on one side of the lattice. Its Hamiltonian can be written as
\begin{equation}
{
 H_{Hei} = -J\cdot \sum_{\langle i,j\rangle}(X_{i}X_{j} + Y_{i}Y_{j} + Z_{i}Z_{j}),
}
\label{eq:MBHVA-Heisenberg}
\end{equation}
where $\langle i,j\rangle$ runs over all the nearest neighbors. Now we consider the construction of CBHVA and MBHVA ans$\rm \ddot{a}$tze.  

According to the rules of HVA mentioned in Section \ref{section:2.1}, given the known form of the Hamiltonian of the Heisenberg model in Eq.~(\ref{eq:MBHVA-Heisenberg}), the construction of a native graph state requires three quantum operations: $R_{XX}$, $R_{YY}$, and $R_{ZZ}$. Owing to the limitation of physical implementation, we restrict the entanglement scheme to $CZ$ under the MBQC to quantitatively discuss the theoretical operation count and verify the feasibility of MBHVA. From Eqs.~(\ref{eq:Rxzx}) and (\ref{eq:Ryzy}), $R_{XX}$ and $R_{YY}$ can be transformed into a combination of single rotation operations and $R_{ZZ}$.  For the construction of the HVA, once the Hamiltonian is known, the required operations for the system are determined. Subsequently, it is possible to obtain the ans$\rm \ddot{a}$tze for both CBQC and MBQC simultaneously, without the need to first construct a quantum circuit and then translate it into the MBHVA. In other words, MBHVA does not rely on a specific form of the quantum circuit and has greater versatility in designing quantum algorithms and optimizing quantum resources.

The quantum circuit of the CBHVA and the graph state of the MBHVA for a system of $N=2 \times 2$ are presented in Fig.~\ref{fig:MBHVA-Heisenberg graph state}. The input state $|\psi_{in}\rangle$ should be initialized to the $|\Psi^{-}\rangle$ Bell State, according to Ref.~\cite{variational-simulation}. In MBHVA, the input state is initialized to the first qubits in the blue nodes, and their measurement patterns are shown in~\ref{appendix:B}.

\subsubsection{Resource costs.}

In this section, we aim to compare the required qubit number and quantum operation count during the execution process of these two different schemes. In our proposed ansatz, the required qubit number, including ancillary qubits, is equal to the operation counts plus the size of the input state while the required qubit number of circuit-based ansatz is fixed according to the problem. Thus, we can focus only on the operation count in the following discussion. The number of measurement operations in the proposed MBHVA is relatively easy to calculate. If the goal of the computation is to obtain the output state, the measurements of the output state can be ignored. In Fig.~\ref{fig:MBHVA-Heisenberg graph state}(b), the blue and yellow nodes require four measurements, the orange nodes require two measurements, and the green nodes require only one measurement. Further details are provided in~\ref{appendix:B}. We consider the operation count of the ansatz without the preparation of the input state because they can be the same for each ansatz. We first calculated the number of measurements required for adjacent particles $2\times 2 (\orangecircle) + 4\times 2 (\bluecircle) + 4\times 2(\yellowcircle) + 3\times 1(\greencircle)=23$ and then multiplied it by the total number of edges $2n(n-1)$ in the two-dimensional lattice. The measurement count for a one-layer MBHVA is $46n(n-1)$. When repeating $D$ times, the total number of measurements will be $46n(n-1)D$.

As shown in Fig.~\ref{fig:MBHVA-Heisenberg graph state}(a), the CBHVA requires only $6n(n-1)D$ gates. We also notice that there are studies on HVA to reduce the resource overhead \cite{HVA_resources1, HVA_resources2} by combining the $R_{XX}$, $R_{YY}$, and $R_{ZZ}$ gates and sharing the same parameters. However, when implementing CBQC ans$\rm \ddot{a}$tze on current quantum hardware, the set of universal quantum gates also needs to be considered, that is, the implementation of multi-qubit rotation gates should be decomposed into combinations of single-qubit rotation gates and two-qubit entangling gates, as shown in Fig.~\ref{fig:multi z}(c), Eqs.~(\ref{eq:Rxzx}) and (\ref{eq:Ryzy}). Taking this decomposition into account, the operation count of CBHVA will be $46n(n-1)D$. The number of single-qubit gates is $34n(n-1)$, and the number of the two-qubit gates is $12n(n-1)$. This is just the same as that of our proposed MBHVA. If the graph state is directly translated from the quantum circuits without optimizing the multi-qubit rotation gate, the required number of measurements can be up to $94n(n-1)D$ for each $CNOT$ gate costs four measurements. From the above discussion, we can see that utilizing the HVA to construct native graph states results in a significant improvement over directly translating quantum circuits in terms of operation counts.

Although the operation count represents only a part of the resource cost, the MBHVA may potentially exhibit significant advantages in certain scenarios. For instance, considering the typical physical realization of CBQC and MBQC, the implementation time of each quantum gate with the superconducting platform is at the nanosecond scale while the execution of each single-qubit measurement only needs time at the picosecond scale with the photonic platform \cite{kjaergaard2020superconducting, barends2014superconducting, walter2017rapid, flamini2018photonic, marsili2013detecting, qiang2018large}. Thus, in this case of the same operation count, MBHVA requires less running time than CBHVA. This advantage in running time means a lower requirement of quantum coherence and less error accumulation. Furthermore, in physical implementations, measurement-independent qubits can be measured simultaneously, this can further enhance the running time advantage. We have provided an explanation of this in~\ref{appendix:C}, and illustrated the correlation between the MBHVA evolution process and physics.

\subsubsection{Evaluation of numerical results.}

In this section, we introduce the V-score method \cite{vscore} to evaluate numerical results. It is particularly suitable for evaluating optimization methods in quantum many-body systems. The definition of the V-score is
\begin{equation}
V\mbox{-}score := \frac{N \cdot {\rm Var}E}{(E-E_{\infty})^{2}},
\label{eq:MBHVA-Vscore}
\end{equation}
where ${\rm Var}E=\langle H^{2} \rangle-\langle H \rangle^{2}$, $E_{\infty}$ serves as a zero point of the energy $E$. It can be observed that the variational optimization method performs better when the V-score is close to zero.

\begin{figure*}[ht]
\centering
\includegraphics[width=0.9\linewidth]{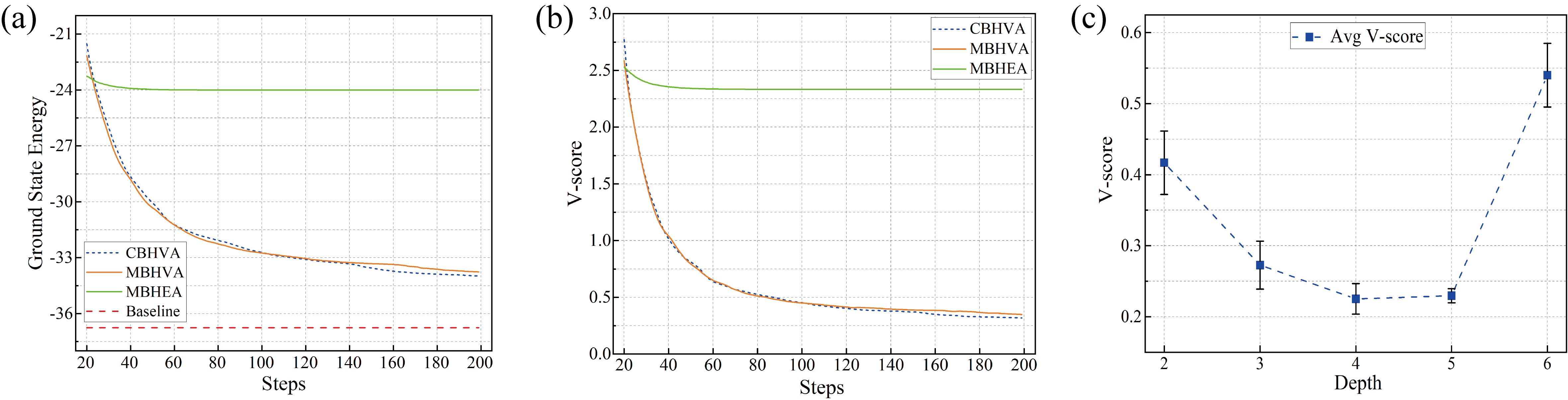}
\caption{\label{fig: Hei-results}(a) The three VQE ans$\rm \ddot{a}$tze results of the ground state energy of the two-dimensional Heisenberg model. The curves are composed of the average of fifty results, and the horizontal axis represents the steps. We limit the steps to 200. (b) By using the V-score as an indicator to evaluate ans$\rm \ddot{a}$tze, the curves are highly similar to those in (a). (c) The V-score of MBHVA varies with different depths. The error bar for each depth is indicated by the variance of ten experiments.}
\end{figure*}

For better illustration, we have set $J=1$, $n=4$, $N=16$, and the open boundary condition. In addition to CBHVA and MBHVA, we also considered the MBVQE proposed in Ref.~\cite{MBVQE} for comparison, whose ansatz is similar to that of hardware efficient ansatz (HEA). Hereafter, we would refer to it as MBHEA. The one-layer building block of the MBHEA is
\begin{equation}
{
 U_{M}(\beta, \gamma) = \left(\prod^{N}_{i=1}R_{X_{i}}(\beta_{1,i})R_{Z_{i}}(\beta_{2,i})R_{X_{i}}(\beta_{3,i})\right)R_{Z^{\otimes N}}(\gamma),
}
\label{eq:MBHVA-MBHEA}
\end{equation}
where $i$ runs over each qubit.

The initial parameters are randomly given. Each ansatz is simulated 50 times, and the average value is taken for each iteration. The depth of each ansatz is set to two. The number of steps is set to 100. Fig.~\ref{fig: Hei-results}(a) and (b) show the numerical results. The exact ground state energy of the system, which is the baseline in Fig.~\ref{fig: Hei-results}(a), is obtained by exact diagonalization (ED). Because we mainly focus on the overall trend during iterations and whether the specific ansatz can effectively approximate the ground state energy, the results of the first 20 steps are not shown in the figure. We have found that the ground state energy and V-score curves of MBHVA and CBHVA are similar, and they can both effectively approach the exact ground state energy. This is expected because their computational processes should both be theoretically efficient. The slight differences are mainly due to the random assignment of the initial parameters. Given the same variational parameters, we have acquired the same output state by using these two ans$\rm \ddot{a}$tze. For the MBHEA, the curve is relatively flat. After the first 20 steps, the ground state energy computed by the MBHEA is reduced to a relatively low value, but the subsequent optimization process may fall into a barren plateau, making it difficult to approach the exact ground state energy. Barren plateau is also a major problem in VQE, and many studies have discussed it \cite{barren_plateaus, barren_plateaus_new, vqa_np, vqa-bp1, vqa-bp2, vqa-bp3, vqa-bp4, vqa-bp5, vqa-bp6}. From the perspective of the evaluation indicator V-score, MBHVA performs better than MBHEA. These results are consistent with the comparison between HVA and HEA in quantum circuits.

In addition to insufficient depth, there are two main reasons why the results of CBHVA and MBHVA deviate slightly from the baseline. First, this deviation can be attributed to the insufficient number of steps, which prevents the results from fully approaching the ground state energy. We set the number of steps to 200 to demonstrate the equivalence of their computations and their superior performance compared with MBHEA. The second reason is that the curve represents the average results. Even if certain experimental results are already sufficiently close to the ground state, the statistical nature of the curve introduces a certain distance from the exact ground state energy. We can conclude that the small deviations observed in the curves of CBHVA and MBHVA from the baseline can be attributed to the limited number of steps and the statistical average of the computed results.

\subsubsection{Performance of variational optimization.}

In addition, the number of variational parameters can also be used as a performance index because it directly affects the efficiency of the classical variational optimization. The total number of parameters for the MBHEA is $(3N+1)D$, which does not change with the observable target. In our experimental setting $(N=16, D=2)$, the number of parameters is 98. The number of parameters for both MBHVA and CBHVA is $6n(n-1)=72$, which can be easily calculated by the number of green nodes in Fig.~\ref{fig:MBHVA-Heisenberg graph state}(b) and \ref{fig:3d-graph}. Because MBHEA has more parameters, it is significantly slower than MBHVA during the variational optimization process. The reason why MBHVA has better optimization performance than MBHEA can also be explained from the perspective of expressivity. Ref.~\cite{Expressibility} proposes a measure of ansatz expressivity, mainly based on the size of the state space covered by the ansatz and the size of the entire Hilbert space of pure states. Ref.~\cite{Expressivity} suggests that for VQA, the best expressivity refers to the state space of the ansatz that precisely covers the optimal point. MBHEA has greater universality, which means that it can access a larger state space and has a larger expressivity and parameter space. However, for the ground state problem of quantum many-body systems, the problem-inspired MBHVA can provide a smaller state space that precisely covers the optimal point, leading to improved performance considering both the ability and efficiency of variational optimization.

Finally, we study the relationship between the depth and performance of MBHVA. We conducted ten experiments respectively with depths varying from one to six and truncated at 100 steps. When $D=1$, MBHVA cannot approach the ground state energy well, with an average ground state energy of about $-25.51$, and a V-score of about $1.81$. When $D\ge 5$, the probability of falling into barren plateaus significantly increases. We have excluded results that fell into barren plateaus for a fair comparison. The results shown in Fig.~\ref{fig: Hei-results}(c) mainly focus on the range $1<D\le 6$. Increasing the number of layers significantly increases the operation count and the number of parameters which would greatly increase the computational load and the probability of the appearance of barren plateaus. In addition, the number of layers is limited by the operation fidelity and decoherence time of the physical hardware. From the experimental results, it can be concluded that selecting an appropriate number of layers is also important for MBVQE.

\subsection{Fermi-Hubbard chain}
\label{section:3.2}

\ \ \ \ The other system used to demonstrate our proposed MBHVA is the one-dimensional Fermi-Hubbard chain, which can describe strong correlation effects in electronic systems, particularly in solid-state materials. The Hamiltonian of the Fermi-Hubbard model is given by
\begin{equation}
{
 H_{Hub} = -t \cdot \sum_{\langle i,j \rangle,\sigma}\left(c^{\dagger}_{i\sigma}c_{j\sigma} + c^{\dagger}_{j\sigma}c_{i\sigma}\right) + U \cdot \sum_{i} n_{i \uparrow}n_{j \downarrow},
}
\label{eq:MBHVA-Hubbard}
\end{equation}
where $t$ is the hopping parameter and $U$ is the Coulomb interaction parameter. We set $t=U=1$.

Once the Hamiltonian is given, we can directly construct the native graph state using HVA. Under the periodic boundary condition, two multi-particle interaction terms will appear which can be represented by the operations $R_{XZ^{\otimes(N-3)}X}$ and $R_{YZ^{\otimes(N-3)}Y}$. When $N>3$, these two operations require $2\times4+1=9$ and $4\times4+1=17$ measurements in the MBQC, while in the CBQC they need $2N+1$ and $2N+9$ gates with the same pattern respectively. Because the number of required operations is fixed in the MBQC, we can conclude that the implementation of these multi-qubit rotation operations is much easier with the MBQC scheme as $N$ grows. If the entanglement scheme is not restricted to the $CZ$ scheme, other multi-qubit rotation operations can also be achieved with only a single measurement, as discussed in~\ref{appendix:A}. This would further enhance the advantages of the MBQC.

\begin{figure*}[ht]
\includegraphics[width=1.0\linewidth]{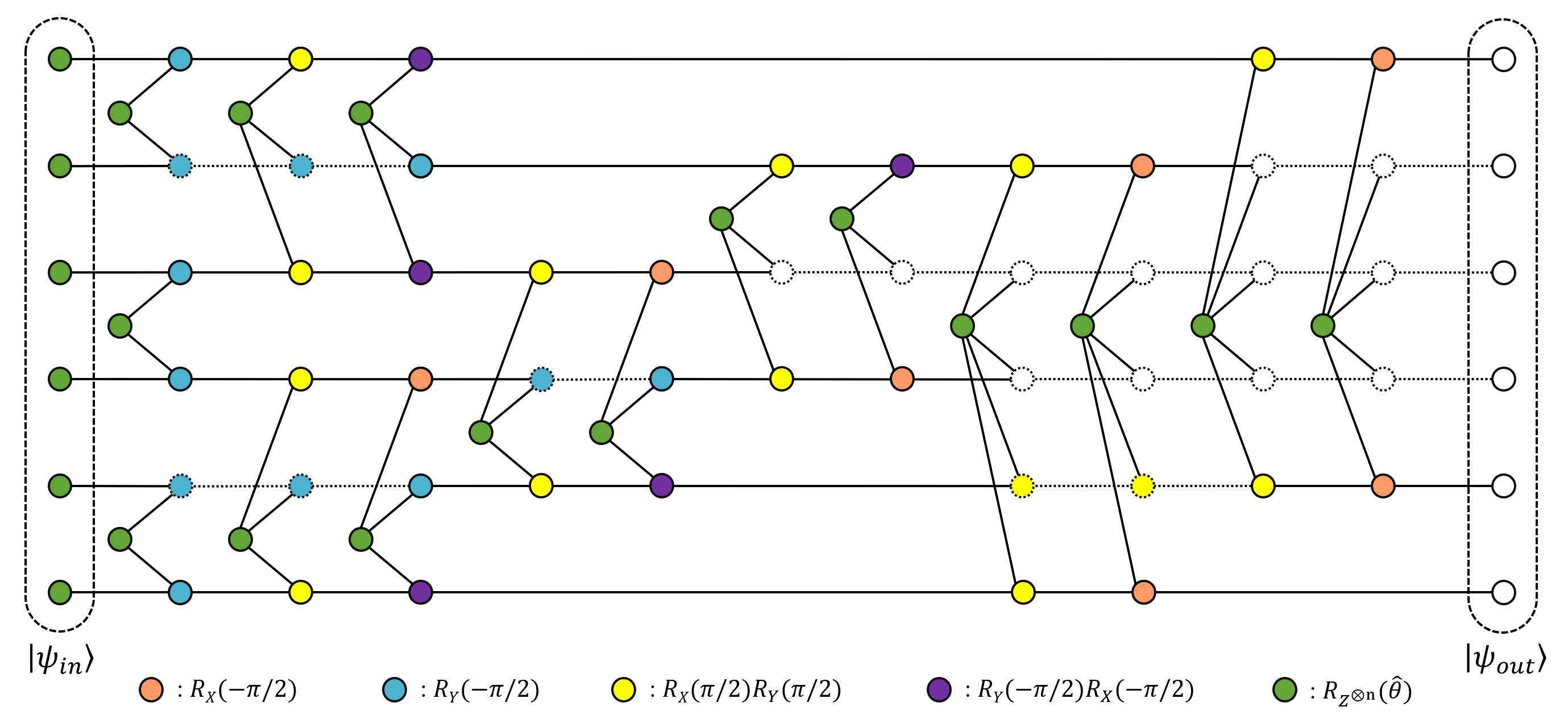}
\caption{\label{fig:MBHVA-hubbard graph state}Schematic representation of the MBHVA graph state for the one-dimensional Fermi-Hubbard chain. Due to the presence of single-qubit, three-qubit, and five-qubit rotation $Z$ operations, we use green nodes to represent the operation $R_{Z^{\otimes n}}$, where $n$ represents the number of non-ancillary qubits involved. The description of the measurement order and phantom nodes are the same as Fig.~\ref{fig:MBHVA-Heisenberg graph state}.}
\end{figure*}

Fig. \ref{fig:MBHVA-hubbard graph state} shows a schematic representation of the MBHVA graph state when particle number $p=3$. According to the Jordan-Wigner (JW) transformation \cite{JW}, it requires $N=6$ qubits. The HVA includes $6$ single-qubit rotation, $3$ two-qubit rotation, $8$ three-qubit rotation, and $4$ five-qubit rotation operations. Thus, in comparison with the Heisenberg model, an additional purple node representing $R_{Y}(-\pi /2)R_{X}(-\pi /2)$ should be introduced. It requires four measurements as a yellow node. According to our schematic representation, it is straightforward to calculate the total number of measurements required for MBHVA. We can simply multiply the measurement count required for each real colored node by its quantity and then sum them to obtain the total measurement count, that is $2\times 7(\orangecircle) + 4\times 7(\bluecircle) + 4\times 12(\yellowcircle) + 4\times 5(\purplecircle) + 21\times 1(\greencircle) = 131$. Similarly, for the CBHVA, we decompose the quantum circuit into elementary single-qubit rotation gates and two-qubit entangling gates, while taking into account the combination of adjacent coaxial rotation gates. After decomposition and optimization, there are 94 single-qubit rotation gates and 70 two-qubit entangling gates, resulting in a total of 164 operations, which is significantly larger than our native MBHVA. Because of the insufficient fidelity of the current physical realizations of two-qubit gates, it is crucial to convert such a large number of two-qubit gates into single-qubit measurement operations within the MBQC scheme. This not only reduces the number of operations but also mitigates error accumulation.

The example of the Fermi-Hubbard chain mentioned here is just one specific case in the context of quantum many-body systems. When the system's Hamiltonian becomes more complicated, the involved multi-qubit rotation operations become more intricate. According to the advantageous characteristics of implementing large multi-qubit rotation operations with the MBQC scheme, MBHVA will have fewer operation counts and effectively mitigate overall error accumulation compared to CBHVA.

\begin{figure}[htb]
\centering
\includegraphics[width=0.5\linewidth]{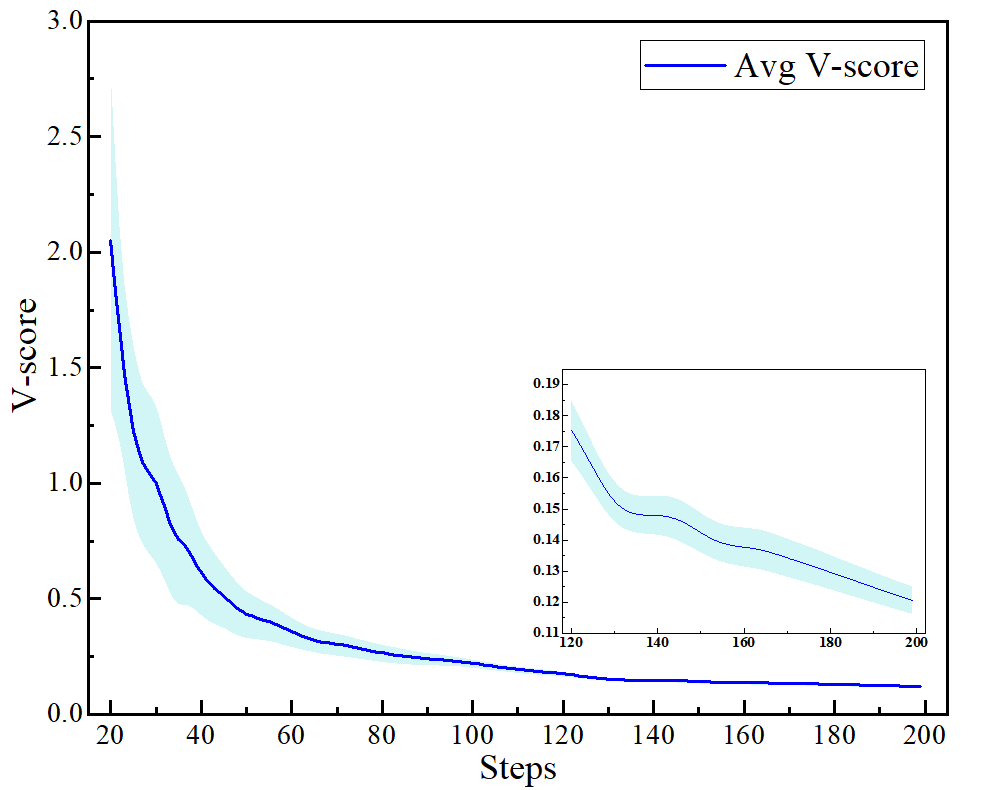}
\caption{\label{fig: Hub-Vscore}V-score varies with steps for finding the ground state of the Fermi-Hubbard chain using MBHVA. The curve has a smooth descent, and the shaded region represents the range of variance. The inset magnifies the interval from step 120 to 200. At step 200, the V-score is sufficiently low, indicating that MBHVA can successfully approach the ground state.}
\end{figure}

We also have calculated the V-score for the case of particle number $p=3$ to demonstrate the variational optimization ability of MBHVA. The initial state is set to $|\Psi^{-}\rangle$, with the initial parameters being randomly assigned. The step size is also set to 200, and the first 20 steps are not shown in the figure. The depth of the ansatz is also set to two. We have found that the ground state energy and V-score curves of MBHVA and CBHVA are similar, and they can both effectively approach the exact ground state energy. Thus, we only depict the curve of MBHVA in Fig.~\ref{fig: Hub-Vscore}. It can be seen that as the number of steps increases, the V-score approaches zero, which indicates that MBHVA can successfully approach the ground state. The average V-score at step $200$ is about $0.12$ and the minimum value is approximately $0.025$. With more steps, the V-score can be even smaller. The size of the shaded area in the figure reflects the range of variance at different steps during the experiments. Smaller shaded areas indicate greater computational stability. This means that MBHVA also has great stability. 

\section{Conclusion and discussion}
\label{section:IV}

\ \ \ \ Considering the fact that the realization of multi-qubit rotation operations, which are very common in the construction of physically-driven VQE, is much easier under the MBQC scheme, we propose the MBHVA method and verify its similar effectiveness and efficiency as CBHVA with the two-dimensional Heisenberg model and demonstrate its advantages in operation count with the Fermi-Hubbard chain. On the whole, there are twofold advantages of our proposed MBHVA. First, with the help of HVA, we can construct the so-called 'native graph state' easily. The resource cost of the native graph state is less than that of the graph state translated from CBQC, thus reducing the need for graph state optimization. Second, in the presence of large multi-qubit rotation operations, the operation count during the execution of the MBHVA is fewer than that of CBHVA, resulting in less running time. This advantage in running time means a lower requirement of quantum coherence and less error accumulation.


It should be emphasized that in our study, we did not consider the impact of errors in either system. Owing to error accumulation, fewer operations may lead to higher computational fidelity and reduce the demand for coherence time. This is an important reason for studying the MBVQE. According to the equivalence of the computation process, commonly used quantum error mitigation techniques \cite{error-mitigation} such as post-selection \cite{post-selection1, post-selection2} and zero-noise extrapolation \cite{ZNE1, ZNE2}, are theoretically applicable to MBVQE as well. Studying error mitigation methods for MBVQE could be a future research prospect \cite{mbqc_fault_tolerance}.

Furthermore, our work is a theoretical analysis based on the $CZ$ entanglement scheme and does not involve specific measurement methods for physical platforms. The actual physical implementation would impose constraints on the physically-driven approach. Fortunately, in complex systems such as the Hubbard model or other more complex models, the physical realization of MBHVA may offer more opportunities for optimization. Therefore, finding the most optimal LC-equivalent MBVQE graph states for non-trivial problems in practice is a complex issue that requires further research.

Last but not least, we discuss the feasibility of actual implementation. From the perspective of classical simulation, MBQC does not require the complete initial preparation of graph states. For the majority of quantum many-body systems, such as the Heisenberg and Hubbard models mentioned in our paper, the worst case only requires generating a graph state with $N+1$ qubits, where $N$ represents the number of qubits associated with the quantum circuit scheme \cite{algorithmic}. This is because we can entangle the qubits required during the execution of the computation. From the perspective of quantum computer experiments, the feasibility of achieving large quantum graph states has been demonstrated in \cite{large_graph_state}. In addition, the progress in single-qubit projection measurement with feed-forward control further strengthens our confidence in achieving near-term quantum advantages using MBVQE.

\section*{Acknowledgement}
We acknowledge the helpful discussions with Zhen Chen at the Beijing Academy of Quantum Information Science regarding the current quantum hardware. This work was supported by the National Natural Science Foundation of China under Grants No.61975005, Beijing Academy of Quantum Information Science under Grants No.Y18G28, and the Fundamental Research Funds for the Central Universities under Grants No.YWF-22-L-938.

\appendix

\section{Measurement patterns for multi-qubit rotation operations}
\label{appendix:A}

In Section \ref{section:2.1}, we mentioned that the representation of the measurement basis takes the form of Eqs.~(\ref{eq:MBQC-basis1}) and (\ref{eq:MBQC-basis2}). When set $\varphi = \pi /2$, the measurement basis becomes

\begin{eqnarray}
|\uparrow(\theta, \pi /2)\rangle &=& {\rm cos}\left(\frac{\theta}{2}\right)|0\rangle+i{\rm sin}\left(\frac{\theta}{2}\right)|1\rangle
, \label{eq:MBQC-basis1-pi2}
\\
|\downarrow(\theta, \pi /2)\rangle &=& -{\rm sin}\left(\frac{\theta}{2}\right)|0\rangle+i{\rm cos}\left(\frac{\theta}{2}\right)|1\rangle
. \label{eq:MBQC-basis2-pi2}
\end{eqnarray}

We donate measurement plane $YZ$ as
\begin{equation}
    \mathcal{M}_{YZ}(\theta):=\{R_{X}(\theta)|0\rangle, R_{X}(\theta)|1\rangle\}
    .
    \label{Myz}
\end{equation}

When apply $X$-axis rotation gates with a rotation angle $-\theta$ to $|0\rangle$ and $|1\rangle$, they are 
\begin{eqnarray}
R_{X}(-\theta)|0\rangle &=& {\rm cos}\left(\frac{\theta}{2}\right)|0\rangle+i{\rm sin}\left(\frac{\theta}{2}\right)|1\rangle
, \label{eq:Myz-basis1-pi2}
\\
R_{X}(-\theta)|1\rangle &=& -{\rm sin}\left(\frac{\theta}{2}\right)|0\rangle+i{\rm cos}\left(\frac{\theta}{2}\right)|1\rangle
. \label{eq:Myz-basis2-pi2}
\end{eqnarray}

The representations of the measurement basis in Eqs.~(\ref{eq:MBQC-basis1-pi2}) \& (\ref{eq:MBQC-basis2-pi2}) and Eqs.~(\ref{eq:Myz-basis1-pi2}) \& (\ref{eq:Myz-basis2-pi2}) are identical, meaning that the measurement basis $\{|\uparrow(\theta, \pi /2)\rangle, |\downarrow(\theta, \pi /2)\rangle\}$ is exactly 
\begin{equation}
    \mathcal{M}_{YZ}(-\theta):=\{R_{X}(-\theta)|0\rangle, R_{X}(-\theta)|1\rangle\}.
\label{eq:Myz-}
\end{equation}

To implement a multi-qubit rotation $Z$ operation, we first define an input state $|\psi_{in}\rangle$ with $N$ qubits. Then, introduce an ancillary qubit, denoted as $a$, which is prepared in the state $|+\rangle_{a}$. According to Eq.~(\ref{eq:MBQC-graph_state}) and the graph structure shown in Fig.~\ref{fig:multi z}, we entangle the ancillary qubit $a$ with each qubit $i\in[N]$ of the input state $|\psi_{in}\rangle$ by applying a $CZ$ operation, where $a$ is the control qubit and $i$ is the target qubit. The corresponding graph state is represented by
\begin{equation}
    |G\rangle=\frac{1}{\sqrt{2}}(|0\rangle_{a}\otimes|\psi_{in}\rangle+|1\rangle_{a}\otimes Z^{\otimes N}|\psi_{in}\rangle)
    .
\end{equation}

Next, we perform a measurement $\mathcal{M}_{YZ}(-\theta)$ using Eq.~(\ref{eq:Myz-}) on qubit $a$. Denote the measurement outcome as $s\in\{0, 1\}$. After the measurement, the output state depends on the measurement result.

When $s = 0$, the output state is
\begin{equation}
    |\psi_{out}\rangle={\rm exp}\left(-i\theta Z^{\otimes N} /2\right)|\psi_{in}\rangle
    .
\end{equation}

When $s = 1$, the output state is
\begin{equation}
    |\psi_{out}\rangle=Z^{\otimes N}{\rm exp}\left(-i\theta Z^{\otimes N} /2\right)|\psi_{in}\rangle
    .
\end{equation}

\begin{figure}[htb]
\centering
\includegraphics[width=0.6\linewidth]{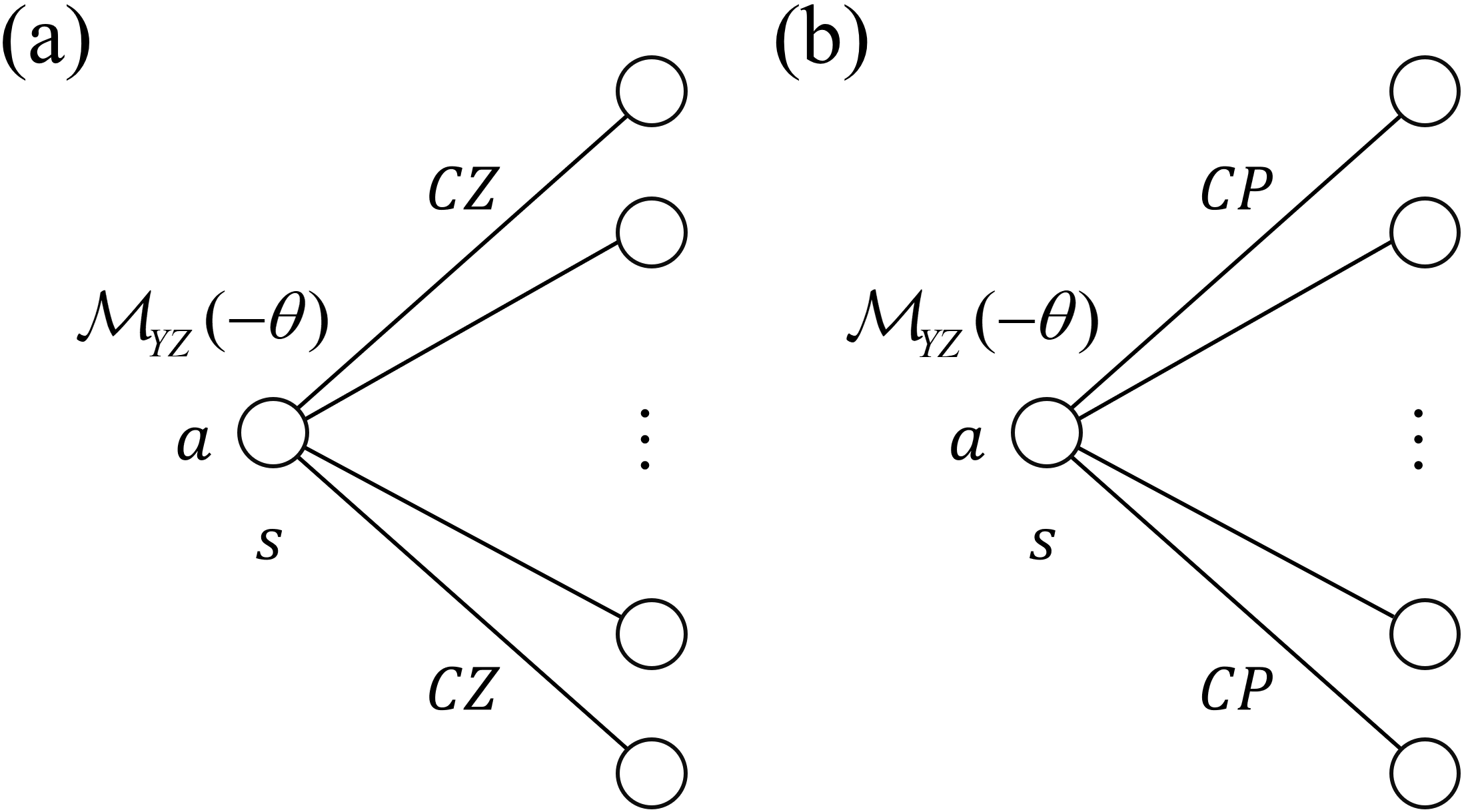}
\caption{\label{fig:multi p}(a) Specific measurement pattern for multi-qubit $Z$ rotation operation where the single measurement acts on ancillary qubit $a$. (b) The extension of (a) by simply modifying the corresponding entanglement scheme from $CZ$ to $CP$.}
\end{figure}

The single measurement described above completes the multi-qubit rotation $Z$ operation, and the measurement pattern is shown in Fig.~\ref{fig:multi p}(a). When the measurement outcome is $1$, it requires the correction of the generated by-product operator $Z^{\otimes N}$. 

If we change the entanglement scheme from $CZ$ into $CX$, then we can realize $R_{X^{\otimes N}}$; into $CY$, $R_{Y^{\otimes N}}$ can be also realized. In conclusion, by applying controlled-$P$ operation as entanglement, where $P \in \{X, Y, Z\}$ is the Pauli operation, the structure of ${\rm exp}\left(-i\theta P^{\otimes N} /2\right)$ as Fig.~\ref{fig:multi p}(b) is realized by a single qubit measurement $\mathcal{M}_{YZ}(-\theta)$ to the ancillary qubit. This demonstrates that MBQC exhibits high flexibility and efficiency in executing multi-qubit rotation operations under arbitrary entanglement schemes.

\section{Measurement patterns and counts for nodes}
\label{appendix:B}

In our paper, we proposed that the node representation is clear and intuitive for translating quantum circuits to graph states. Additionally, once the measurement count required for each node is known, calculating the total measurement count becomes straightforward. However, it is important to know the operations represented by each node and their corresponding measurement patterns to construct a complete MBQC graph state. Before discussing the measurement patterns and counts for each node, we first introduce the following basic formulas:
\begin{eqnarray}
    HXH&=&Z
    , \label{eq:hxh}\\
    HZH&=&X
    , \label{eq:hzh}\\
    \{X, Z\}&=&0
    . \label{eq:XZ}
\end{eqnarray}

These formulas demonstrate the conversion relationship between the Pauli $X$ and Pauli $Z$ operators, which are anti-commutative.

Our discussion is based on the $CZ$ entanglement scheme. As for the measurement pattern of the green nodes, it has already been indicated in~\ref{appendix:A} that it only requires a single measurement on the ancillary qubit. 

For the other colored nodes, we need to introduce the measurement plane $XY$ as
\begin{equation}
    \mathcal{M}_{XY}(-\theta):=\{R_{Z}(-\theta)|+\rangle, R_{Z}(-\theta)|-\rangle\}
    .
    \label{Myz}
\end{equation}

\begin{figure}[htb]
\centering
\includegraphics[width=0.7\linewidth]{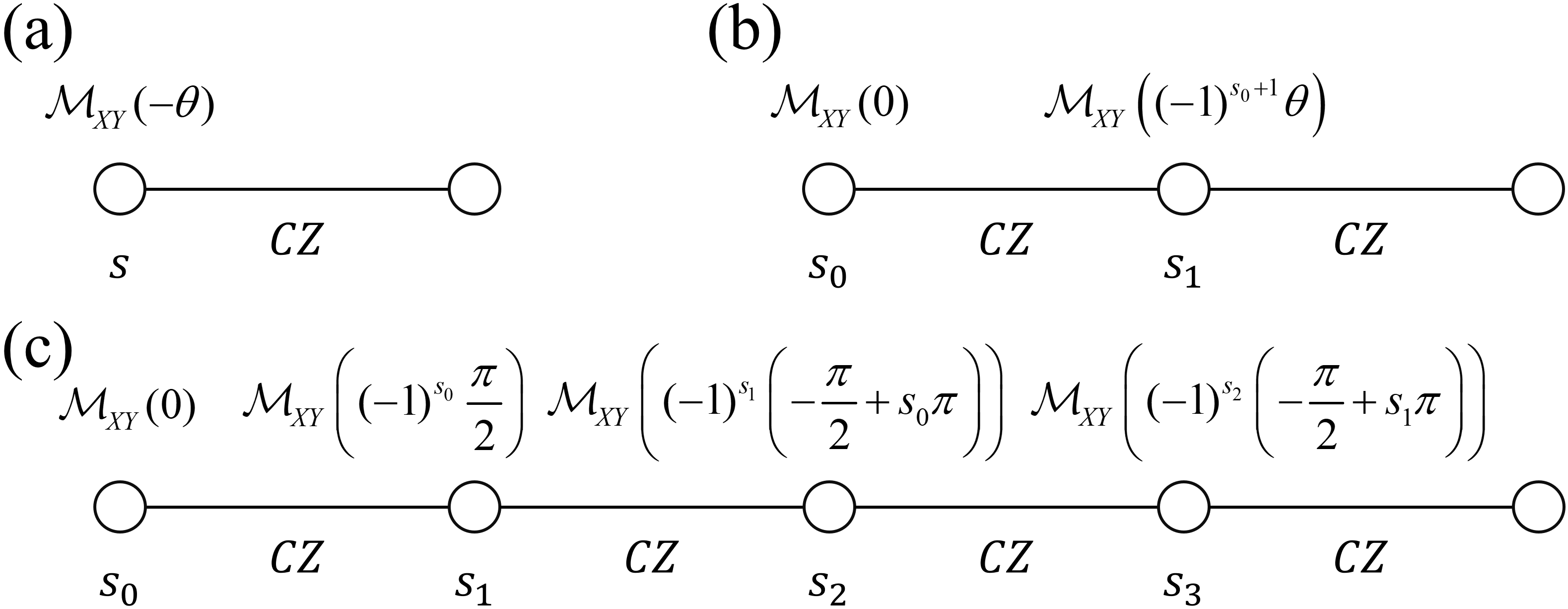}
\caption{\label{fig:patterns}(a) Measurement pattern for $R_{X}(\theta)H$. Only a single measurement is needed. (b) Measurement pattern for $R_{X}(\theta)$. The measurement order is from left to right. (c) Measurement pattern for $R_{Y}(-\pi/2)$. The operation can be converted to the combination of the former two, thus it requires four measurements.}
\end{figure}

First, assuming that the desired operation to be performed is $R_{X}(\theta)H$, the output state is
\begin{equation}
    |\psi_{out}\rangle=R_{X}(\theta)H|\psi_{in}\rangle
    .
\end{equation}

Constructing a measurement pattern as shown in Fig.~\ref{fig:patterns}(a), after the measurement is completed, the output state is
\begin{equation}
    |\psi_{out}\rangle=X^{s}R_{X}(\theta)H|\psi_{in}\rangle
    , \label{eq:mbqc-RxH}
\end{equation}
where $X^{s}$ represents the by-product operator and is determined by the outcome of the measurement. Note that the aforementioned operation can be completed via a single measurement. To obtain the measurement pattern for $R_{X}(\theta)$, we must remove $H$, which requires an additional measurement operation. Therefore, the measurement pattern for $R_{X}(\theta)$ is depicted in Fig.~\ref{fig:patterns}(b), and can be prepared as:
\begin{equation}
    |\psi_{out}\rangle=X^{s_{1}}R_{X}\left((-1)^{s_{0}}\theta\right)HX^{s_{0}}H|\psi_{in}\rangle
    , \label{eq:mbqc-Rx}
\end{equation}
which is equal to
\begin{equation}
    |\psi_{out}\rangle=X^{s_{1}}Z^{s_{0}}R_{X}(\theta)|\psi_{in}\rangle
    ,
\end{equation}
that is to say that the operation $R_{X}(-\pi/2)$ represented by orange node should cost two measurements.

Next, we focus on the blue nodes that represent $R_{Y}(-\pi/2)$. Operations $R_{Y}(-\pi/2)$ and $R_{Y}(\pi/2)$ are both involved in the graph state and can be discussed together. Owing to the limitations of the entanglement scheme, we need to decompose them further:
\begin{eqnarray}
    R_{Y}\left(-\frac{\pi}{2}\right)=R_{X}\left(\frac{\pi}{2}\right)R_{Z}\left(\frac{\pi}{2}\right)R_{X}\left(-\frac{\pi}{2}\right)
    , \label{eq:Ry-}\\
    R_{Y}\left(\frac{\pi}{2}\right)=R_{X}\left(\frac{\pi}{2}\right)R_{Z}\left(-\frac{\pi}{2}\right)R_{X}\left(-\frac{\pi}{2}\right)
    . \label{eq:Ry}\\ \nonumber
\end{eqnarray}

By utilizing Eq.~(\ref{eq:hxh}), we can transform $R_{Y}(-\pi/2)$ and $R_{Y}(\pi/2)$ as
\begin{eqnarray}
    R_{Y}\left(-\frac{\pi}{2}\right)=R_{X}\left(\frac{\pi}{2}\right)HR_{X}\left(\frac{\pi}{2}\right)HR_{X}\left(-\frac{\pi}{2}\right)
    , \label{eq:Ry-H}\\
    R_{Y}\left(\frac{\pi}{2}\right)=R_{X}\left(\frac{\pi}{2}\right)HR_{X}\left(-\frac{\pi}{2}\right)HR_{X}\left(-\frac{\pi}{2}\right)
    . \label{eq:Ry+H}
\end{eqnarray}

This form is well-suited for constructing the measurement pattern using Eqs.~(\ref{eq:mbqc-RxH}) and (\ref{eq:mbqc-Rx}). Therefore, the blue node, as shown in Fig.~\ref{fig:patterns}(c) requires four measurements, and the measurement pattern takes the form:
\begin{equation}
    \begin{split}
    |\psi_{out}\rangle= & X^{s_{3}}R_{X}\left((-1)^{s_{2}}\left(\frac{\pi}{2}-s_{1}\pi\right)\right)HX^{s_{2}}R_{X}\left((-1)^{s_{1}}\left(\frac{\pi}{2}-s_{0}\pi\right)\right)\\
    & HX^{s_{1}}R_{X}\left((-1)^{s_{0}}\left(-\frac{\pi}{2}\right)\right)HX^{s_{0}}H|\psi_{in}\rangle
    .
    \end{split}
\end{equation}

According to Eqs.~(\ref{eq:hxh}), (\ref{eq:hzh}) and (\ref{eq:XZ}), after performing algebraic operations, the output state will be
\begin{equation}
    |\psi_{out}\rangle=Z^{s_{2}}X^{s_{3}}R_{X}\left(\frac{\pi}{2}\right)HR_{X}\left(\frac{\pi}{2}\right)HR_{X}\left(-\frac{\pi}{2}\right)|\psi_{in}\rangle
    ,
\end{equation}
means the measurement pattern executes operations equivalent to $R_{Y}(-\pi/2)$.

Finally, we discuss the yellow and red nodes, which involve consecutive applications of $R_{Y}$ and $R_{X}$ operations. By employing Eqs.~(\ref{eq:Ry-H}) and (\ref{eq:Ry+H}), we can expand their operations as
\begin{equation}
    R_{Y}\left(-\frac{\pi}{2}\right)R_{X}\left(-\frac{\pi}{2}\right)=R_{X}\left(\frac{\pi}{2}\right)HR_{X}\left(\frac{\pi}{2}\right)HR_{X}(-\pi)
    , \label{eq:RyRx}
\end{equation}
\begin{equation}
    R_{X}\left(\frac{\pi}{2}\right)R_{Y}\left(\frac{\pi}{2}\right)=R_{X}(\pi)HR_{X}\left(\frac{\pi}{2}\right)HR_{X}\left(-\frac{\pi}{2}\right)
    . \label{eq:RxRy}
\end{equation}

When two consecutive $R_{X}$ are adjacent, their angles can be combined, thereby reducing the number of measurement operations. This holds true for quantum circuits as well, where it can reduce the number of quantum gate operations. By observing the form of Eqs.~(\ref{eq:RyRx}) and (\ref{eq:RxRy}), they are very similar to Eqs.~(\ref{eq:Ry-H}) and (\ref{eq:Ry+H}). $R_{X}$ with angle $\pi$ or $-\pi$ does not reduce to $I$, but can be represented by Pauli $X$. Therefore, the yellow and red nodes also require four measurements as the blue nodes, and their measurement patterns closely resemble the one shown in Fig.~\ref{fig:patterns}(c).

\section{Physically-driven measurement-based evolution process}
\label{appendix:C}

\begin{figure}[h]
\centering
\includegraphics[width=0.6\linewidth]{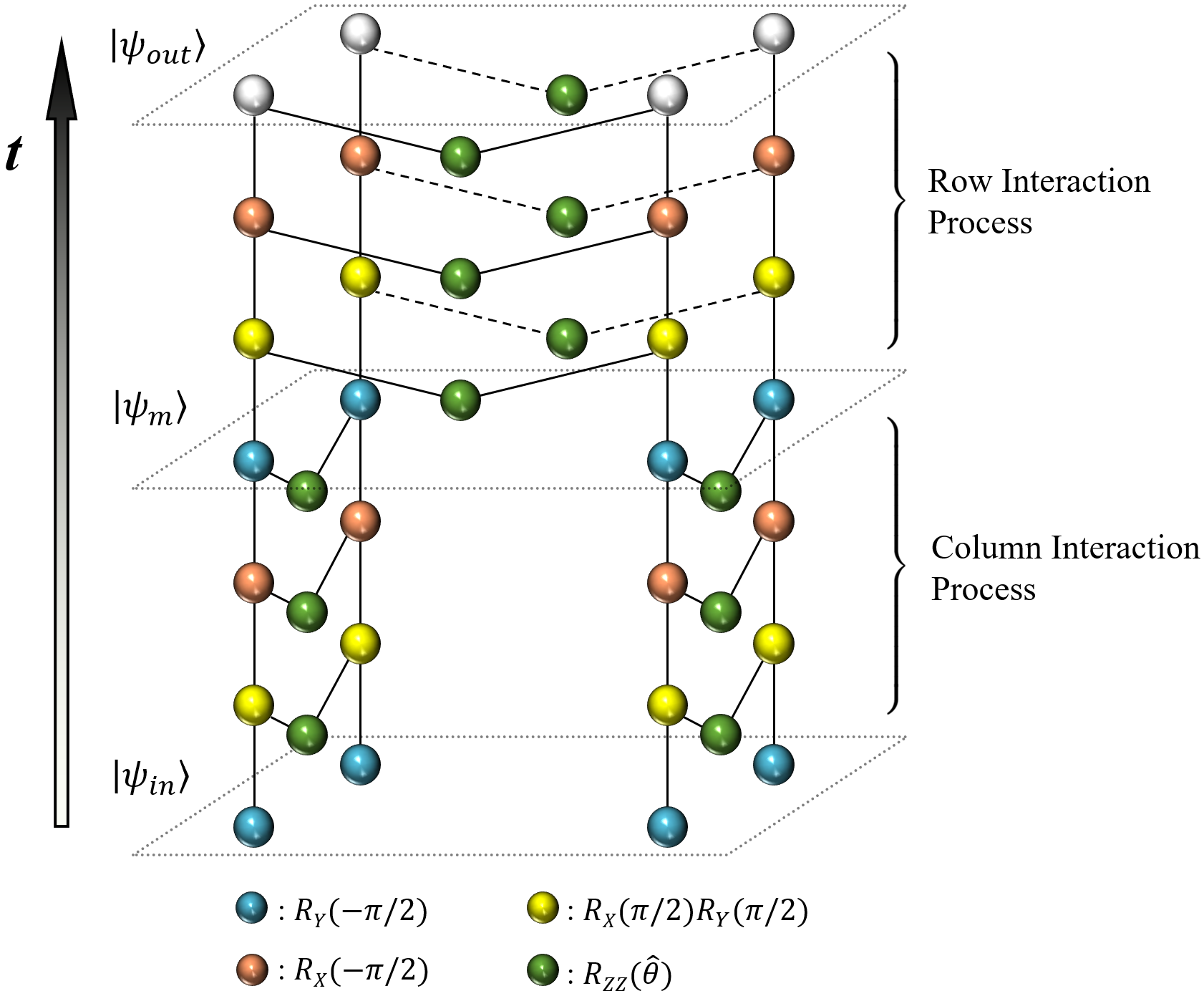}
\caption{\label{fig:3d-graph}A three-dimensional schematic representation of the one-layer graph state. The interaction terms are effectively depicted on the nodes in the figure.}
\end{figure}

It is worth pointing out that the obtained graph state of the MBHVA can also be understood from the perspective of three dimensions. As shown in Fig.~\ref{fig:3d-graph}, the evolution of the graph state through the measurements is from the bottom to the top over time. The three dashed-line planes represent three state stages. The first state stage represents the initial state, followed by evolution via column interactions to reach the middle stage. Finally, the output state is obtained through evolution via row interactions. Nodes without dependencies, such as those in the same horizontal plane, can simultaneously execute their measurement patterns. In physically implementation, this can reduce the running time of MBHVA further. The construction of the graph state and the sequence of measurements are based on the form of the Hamiltonian, which interprets the physically-driven measurement-based evolution process.

\section*{References}
\bibliography{iop_AMBQCPDVQE.bib}

\end{document}